\documentclass[prd,preprintnumbers,amsmath,amssymb,twocolumn,nofootinbib,longbibliography]{revtex4-1}
\usepackage[utf8]{inputenc}

\usepackage{multirow}
\usepackage{graphicx}
\usepackage{booktabs}
\usepackage[hyperfootnotes=false]{hyperref}
\usepackage{color}
\usepackage{slashed}
\usepackage{epstopdf}
\usepackage{cancel} 

\newcommand{\bea}{\begin{eqnarray}}
\newcommand{\eea}{\end{eqnarray}}

\definecolor{cBlue}{RGB}{0,110,191}
\definecolor{cLightBlue}{RGB}{214,237,252}
\definecolor{cRed}{RGB}{196,0,100}
\definecolor{cLightRed}{RGB}{254,222,237}
\definecolor{cGreen}{RGB}{0,166,80}

\DeclareUnicodeCharacter{2212}{-}
\DeclareUnicodeCharacter{2217}{*}

\begin{document}

\title{A novel determination of the $B_c$ lifetime}

\author{Jason Aebischer}
\email{jaebischer@physics.ucsd.edu}
\author{Benjam\'in Grinstein}

\email{bgrinstein@ucsd.edu}
\affiliation{Department of Physics, University of California at San Diego, La Jolla, CA 92093, USA}

\begin{abstract}
To reduce the current theory uncertainties  a novel way to determine  the $B_c$ lifetime, $\tau_{B_c}$,  is proposed. Taking the difference of the $B_c$, and  $B$ and $D$ meson decay rates eliminates the leading contributions from the calculation, which exhibit large scale and scheme dependence. The uncertainties in the proposed determination of $\tau_{B_c}$ are analyzed and improvements are proposed. The method predicts a value of $\tau_{B_c}$ in tension with the experimental determination. Several explanations are considered,  including underestimation of uncertainties, duality violation and new physics. We discuss quantitative evidence that our method is vitiated by  duality violation in the large mass OPE used to calculate nonleptonic decay rates of $B$ and $D$ mesons.

\end{abstract}

\maketitle

\section{Introduction}
Recently, the Standard Model (SM) prediction of the $B_c$ lifetime has been updated in \cite{Aebischer:2021ilm}, using an Operator Product Expansion (OPE). The results presented in three different mass schemes exhibit however a large scheme dependence, which does not allow for a clear-cut conclusion concerning New Physics effects in the $B_c$ decay. In this work we propose an alternative approach to estimate the $B_c$ decay width, which reduces the level of scheme dependence.

The idea is to relate the decay rate of the $B_c$ to the ones from the $B$ the $D$ mesons. Assuming quark-hadron duality a heavy quark expansion can be formulated so that the decay rate of a meson $H_Q$ with heavy quark $Q$ can be written schematically as
\begin{equation}
\label{eq:preT_M}
    \Gamma(H_Q) = \Gamma_Q^{(0)}\left(C_0 +\sum_{n\ge2} \frac{C_n}{m_Q^n}\langle\mathcal{O}^{(n)}\rangle\right)\,,
\end{equation}
where $C_n$ can be computed perturbatively and the generally non-perturbative expectation value of the operators $\mathcal{O}^{(n)}$ is computed in the single particle heavy meson state $H_Q$.  The coefficient $C_0$ is implicitly multiplied by  the exactly known expectation value of the $Q$-number density operator, so that the iso-singlet $\Gamma^{\rm pert}_{\rm sing}=\Gamma^{(0)}_Q C_0$ is precisely the decay rate of the unbound heavy quark. Again assuming quark-hadron duality, a similar expansion can be formulated for quarkonium (heavy-heavy mesons), now based on a non-relativistic expansion (NRQCD). In this case the expansion of the decay rate has formally the same form as in Eq.\eqref{eq:preT_M}, and in particular the leading term is given by the perturbative one. Since the $B$ and $D$ mesons decay by $b$-quark or $c$-quark decay, respectively, and the  $B_c$ may decay by either of these, using Eq.~\eqref{eq:preT_M} immediately gives
\begin{multline}
\label{eq:relBc}
\Gamma(B)+\Gamma(D)-\Gamma(B_c) = \Gamma_{\rm sing}^{n.p.}(B)+\Gamma_{\rm sing}^{n.p.}(D)-\Gamma_{\rm sing}^{n.p.}(B_c)\\
+\Gamma^{\text{WA}+\text{PI}}(B)+\Gamma^{\text{WA}+\text{PI}}(D)-\Gamma^{\text{WA}+\text{PI}}(B_c)\,.
\end{multline}
Irrespective of the fact that the contributions from the $B$ and $D$ mesons are computed in Heavy Quark Effective Theory (HQET) whereas for $B_c$ Non-Relativistic QCD (NRQCD) is employed, the free quark decay rates cancel in the difference of Eq.~(\ref{eq:relBc}). This is a consequence of quark number symmetry, which allows to normalize the matrix elements of the leading operators to one, irrespective of the flavour content.
The left hand side fails to vanish by the non-perturbative (n.p.) corrections  to the quark decay rate, $\Gamma^{n.p.}_{\rm sing}$,  as well as, depending on the charge of the meson, by the Weak Annihilation (WA) and Pauli-Interference (PI) contributions, $\Gamma^{\text{WA}}$ and $\Gamma^{\text{PI}}$. The right hand side (RHS) of Eq.~\eqref{eq:relBc} is not free from the perturbative uncertainties that are evident in the quark decay rate when different mass schemes are compared, but it is {\it a priori} smaller, so that the largest uncertainty in the calculations of the width is eliminated.

 Following Eq.~(\ref{eq:relBc}), the $B_c$ decay rate can therefore be expressed in terms of the $B$ and $D$ decay rates as well as their n.p. corrections and WA/PI contributions. For $\Gamma(B)$ and $\Gamma(D)$ the experimentally measured values given in Tab.~\ref{tab:num-input} will be used, which leaves a computation of the n.p. and WA/PI contributions. We will present the calculation of these terms in the next section. Then in Sec.~\ref{Sec:Discussion} we look closely and critically at the ingredients of the calculation. We end by offering take-aways in Sec.~\ref{Sec:Conclusions}.

\section{Computation}

For the computation of the n.p.~$B_c$ contributions we follow closely \cite{Aebischer:2021ilm}, where the non-perturbative velocity expansion was carried out up to third order in the relative velocity of the heavy quarks. Employing NRQCD, we keep terms consistently up to second order in the relative velocity expansion, which consists of the kinetic energy term \begin{equation}\label{eq:kinEn}
   \Gamma_{\rm sing}^{n.p.}(B_c)= \Gamma^{(0)}_{Q\to q'}\tfrac1{2m_Q^2}\langle B_c|\overline\Psi_+ (iD_\perp)^2 \Psi_+|B_c\rangle \,,
\end{equation}
depending on the positive energy component spinor $\Psi_+$, satisfying $\Psi_+=\left(\frac{1+\slashed{u}}{2}\right)\Psi_+$ with the center of mass 4-velocity $u$. In Eq.~\eqref{eq:kinEn} a summation over both  $Q=\bar b$ and $Q=c$ quarks is implied, with the leading order (LO) decay rate $\Gamma^{(0)}_{Q\to q'}=\frac{G_F^2m_Q^5}{192\pi^3}|V_{Qq'}|^2$. Higher order corrections to this expression result from $\alpha_s$-corrections to the kinetic term and the decay rate, as well as from higher order operators in the velocity expansion. We use here the contributions from the chromomagnetic operator and the Darwin term as an estimate for higher velocity corrections to the leading term in Eq.~(\ref{eq:kinEn}). Furthermore, the WA/PI contributions of the $B_c$ are also taken into account because of their phase space enhancement factor of $16\pi^2$, even though they are formally of higher order in the velocity expansion. As discussed in \cite{Aebischer:2021ilm}, spin-symmetry is used to reduce the number of bag parameters in the computation of
  WA and PI contributions to the $B_c$ width.

For the $H_Q=B,D$ mesons, using HQET the n.p. corrections read
\begin{multline}
\label{eq:T_M}
    \Gamma^{n.p}(H_Q) = \Gamma_Q^{(0)}\left(C_{\lambda_1}\left(\frac{\lambda_1}{2m_Q^2}+\frac{\tau_1+3\tau_2}{2m_Q^3}\right)\right.\\+C_{\lambda_2}\left(\frac{3\lambda_2}{2m_Q^2}+\frac{\tau_3}{2m_Q^3}\right)+
    \left.\sum_{i=1}^2C_{\rho_i}\frac{\rho_i}{2m_Q^3}\right)\\
    +\sum_{i,q}C_{4F_i}^{(q)}\frac{\langle \mathcal{O}_{4F_i}^{(q)}\rangle}{4m_Q^3}+\mathcal{O}(1/m_Q^4)\,,
\end{multline}
where the free quark decay rate is given by $\Gamma_Q^{(0)}C_{\lambda_1}$.
  The matrix elements in Eq.~(\ref{eq:T_M}) are defined as:
\begin{align}
\label{eq:lambda1defd}
    &\langle H_Q|\bar h_Q (i(D_\perp)_\mu)(iD_\perp^\mu)h_Q|H_Q\rangle = \lambda_1 +\frac{\tau_1+3\tau_2}{m_Q}\,,\\
\label{eq:lambda2defd}
&\langle H_Q|\bar h_Q\frac{g}{2}\sigma_{\mu\nu}G_\perp^{\mu\nu}h_Q|H_Q\rangle = 3\lambda_2+\frac{\tau_3}{m_Q} \,,\\
\label{eq:rho1defd}
&\langle H_Q|\bar h_Q (i(D_\perp)_\mu)(iv\cdot D_\perp)(iD_\perp^\mu) h_Q|H_Q\rangle = \rho_1 \,,\\
\label{eq:rho2defd}
&\langle H_Q|\bar h_Q (iD_\perp^\mu)(iv\cdot D_\perp)(iD_\perp^\nu)(-i\sigma_{\mu\nu}) h_Q|H_Q\rangle = \rho_2 \,,
\end{align}
where $H_Q$ denotes a HQET state characterized by 4-velocity $u$, and $D_\perp^\mu=D^\mu-(u\cdot D)u^\mu$.
The coefficients $C_{\lambda_1}$ correspond to the perturbative corrections to the free quark decay rate and are computed in \cite{Guberina:1979fe,Altarelli:1980fi,Altarelli:1991dx,Buchalla:1992gc,Ho-kim:1983klw}. $C_{\lambda_2}$ is computed at tree level in \cite{Manohar:1993qn} for semileptonic decays, for non-vanishing lepton masses in \cite{Falk:1994gw} and for hadronic decays in \cite{Bigi:1992su,Blok:1992hw,Blok:1992he,Bigi:1992ne}. The coefficients $C_{\rho_{1,2}}$ are computed in \cite{Gremm:1996df} for semileptonic decays, and are  inferred from
the contributions from the Darwin and spin-orbit terms computed recently for hadronic decays in \cite{Mannel:2020fts,Lenz:2020oce}.
On the right hand side of Eq.~\eqref{eq:T_M} matrix elements    have been parametrized by the mass-independent HQET matrix elements of local operators $\lambda_{1,2}$, $\rho_{1,2}$ as well the time-ordered products $\tau_{1,2,3}$ \cite{Gremm:1996df} defined in Eqs.~\eqref{eq:lambda1defd}--\eqref{eq:rho2defd}. The results in Refs.~\cite{Mannel:2020fts,Lenz:2020oce} are presented in terms of the set of mass-dependent parameters $\mu_\pi$, $\mu_G$, $\rho_D$, $\rho_{LS}$, which can not be used for the $B$ and $D$ decays simultaneously, due to their quark mass dependence; Ref.~\cite{Gremm:1996df} explains how to  translate between these two parametrizations.
For the parameters numerical values we use the most recent fit results from HFLAV \cite{HFLAV:2019otj},%
\footnote{Ref.~\cite{Lenz:2020oce} differs in its choice of parametrization from Refs.~\cite{Gremm:1996df,Mannel:2020fts} in that their operators involve $D^\mu$ rather than $D_{\perp}^\mu$, that is, they do not project out the component of the covariant derivative that is perpendicular to the 4-velocity of the heavy quark. The translation  between these bases is given in~\cite{Dassinger:2006md}. HFLAV does not explicitly state which of these bases is used in its fit; we assume their choice corresponds to that of Refs.~\cite{Gremm:1996df,Mannel:2020fts} }
see Tab.~\ref{tab:num-input}.
Furthermore, to be consistent with our scheme choice, we use HQET expansions of meson masses to fix the parameters $\lambda_2$ and $\rho_2$ from $m_B-m_{B^*}$ and $m_D-m_{D^*}$, and $m_c$ from   $m_B-m_D$ and $m_b$ determined in the 1S expansion \cite{Hoang:1998hm,Hoang:1998ng,Hoang:1998nz}. This choice of heavy quark masses corresponds to what is referred to as the ``meson scheme" for quark masses in \cite{Aebischer:2021ilm}.

We approximate $u$, $d$ and $s$ quarks as massless. The calculation of the  Darwin term's contribution to the decay rate of $B$ and $D$ mesons exhibits an infrared divergence  for all massless final state quarks. We use the result in Ref.~\cite{Lenz:2020oce} which subtracts the IR divergence in the coefficient of the Darwin term against the divergence in the 4-quark operator; we convert the result to the basis of operators used by HFLAG (also by \cite{Gremm:1996df} and \cite{Mannel:2020fts}). This means that when the ``eye-graph" is computed on the lattice an explicit subtraction of the IR divergence is needed.\footnote{For a definition of the ``eye-graphs" see graphs (b) in Fig.~3 of Ref.~\cite{Becirevic:2001fy}.}

\begin{table}
\centering
\renewcommand{\arraystretch}{1.3}
\resizebox{\columnwidth}{!}{
\begin{tabular}{|llllll|}
\hline
  Parameter
& Value
& Ref.
&  Parameter
& Value
& Ref.
\\
\hline\hline
  $\Gamma_{B^+}$                   & $0.611(1)\, \text{ps}^{-1}$  & \cite{Tanabashi:2018oca}
& $\Gamma_{B^0}$ &  $0.658(2)\, \text{ps}^{-1}$  & \cite{Tanabashi:2018oca}
\\
$\Gamma_{D^+}$                   & $0.962(6)\, \text{ps}^{-1}$  & \cite{Tanabashi:2018oca}
& $\Gamma_{D^0}$ &  $2.438(9)\, \text{ps}^{-1}$  & \cite{Tanabashi:2018oca}
\\
\hline
  $M_{B_c}$     & $6274.9\pm 0.8$ MeV  & \cite{Tanabashi:2018oca}
& $f_{B_c}$                   & $0.427(6)$ GeV         & \cite{McNeile:2012qf}
\\
    $f_D$   &  $0.2090(24)$ GeV & \cite{FlavourLatticeAveragingGroup:2019iem}
& $f_{B}$                   & $0.1920(43)$ GeV         & \cite{FlavourLatticeAveragingGroup:2019iem}
\\
  $M_{B_c^*}-M_{B_c}$                 & $54(3)$ MeV     & \cite{Dowdall:2012ab}
&  $M_{\Upsilon}$             & $9460.30(26)$ MeV           & \cite{Tanabashi:2018oca}
\\
  $M_{B^0}$             & $5279.65(12)$ MeV           & \cite{Tanabashi:2018oca}
& $M_{B^+}$             & $5279.34(12)$ MeV           & \cite{Tanabashi:2018oca}
\\
  $M_{D^0}$             & $1864.84(5)$ MeV           & \cite{Tanabashi:2018oca}
& $M_{D^+}$             & $1869.66(5)$ MeV           & \cite{Tanabashi:2018oca}
\\
 $M_{D^{*0}}$             & $2006.85(5)$ MeV           & \cite{Tanabashi:2018oca}
& $M_{D^{*+}}$             & $2010.26(5)$ MeV           & \cite{Tanabashi:2018oca}
\\
\hline
 $\lambda_1$            & $-0.362(67)$ GeV$^2$           & \cite{HFLAV:2019otj}
& $\rho_1$             & $0.043(48)$ GeV$^3$           & \cite{HFLAV:2019otj}
\\
 $\tau_1$             & $0.161(122)$ GeV$^3$           & \cite{HFLAV:2019otj}
& $\tau_2$             & $-0.017(62)$ GeV$^3$           & \cite{HFLAV:2019otj}
\\
 $\tau_3$             & $0.213(102)$ GeV$^3$           & \cite{HFLAV:2019otj}
&             &           &
\\
\hline
 $B_1(m_b)$             & $1.10(13)\left(\substack{+0.10\\-0.21}\right)$            & \cite{Becirevic:2001fy}
& $ B_2(m_b)$             & $0.79(5)(9)$            & \cite{Becirevic:2001fy}
\\
 $\epsilon_1(m_b)$             & $-0.02(2)\left(\substack{+0.01\\-0.00}\right)$            & \cite{Becirevic:2001fy}
& $  \epsilon_2(m_b)$             & $0.03(1)\left(\substack{+0.01\\-0.00}\right)$            & \cite{Becirevic:2001fy}
\\
\hline
\end{tabular}
}
\caption{\small
Input parameters used for the numerical analysis.
}
  \label{tab:num-input}
\end{table}

In the language of the heavy quark operator expansion, the computation of WA and PI contributions to  decay widths is contained in four fermion operators and their coefficients, {\it e.g.}, $\mathcal{O}^{(q)}_{4Fi}$ and $C^{(q)}_{4Fi}$ in Eq.~\eqref{eq:T_M} for the case of $B$ and $D$ decays.
For the coefficient functions in the OPE for $B$ and $D$ decays we use the NLO calculation of Ciuchini {\it et al} \cite{Ciuchini:2001vx}. The matrix elements are expressed in terms of ``bag'' parameters that are obtained from  Monte Carlo simulations of QCD on the lattice for the case of $B$ decays~\cite{Becirevic:2001fy}. Tab.~\ref{tab:becirevicBagD} gives the $D$-decays  bag parameters at $\mu=2.7(1)$~GeV that we have  read off from Fig.~4 in Ref.~\cite{Becirevic:2001fy}; the coefficient functions were computed at this renormalization point. Note however that the calculation of bag parameters in \cite{Becirevic:2001fy} neglects the contribution of eye graphs.  For $B_c$ decays the contributions of WA and PI are calculated at LO. For $D$ decays NLO QCD corrections to the PI contributions can be large as discussed in \cite{King:2021xqp}.  One should  bear in mind that while both PI and WA graphs contribute to the $B_c$ width, only the WA graphs contribute in the relation of Eq.~(\ref{eq:relBc}) for the neutral modes ({\em i.e.},  $B^0$ and $D^0$), and only PI graphs contribute to the relation for charged modes ({\em i.e.}, $B^+$ and $D^+$).

\begin{table}[h!]
\centering
 \begin{tabular}{|clcl|}
 \hline
 Parameter & Value& Parameter& Value\\ [0.5ex]
 \hline
\hline
$B_1(\mu)$& $0.94(5)$    & $\overline{B}_2(\mu)$& $0.46(2) $\\
\hline
$\epsilon_1(\mu)$ &  $-0.051(15)$   & $\overline{\epsilon}_2(\mu)$ &$0.005(3)$ \\[0.5ex]
 \hline
 \end{tabular}
 \caption{\small
Bag parameters at $\mu=2.7(1)$~GeV used in the computation of $D$-meson WA and PI contributions to the decay width. Inferred from Fig.~4 in Ref.~\cite{Becirevic:2001fy}. The barred symbols are defined through $X=m_D^2/(m_c+m_q)^2\overline{X}$.
}
  \label{tab:becirevicBagD}
\end{table}

\begin{table}[h!]
\centering
 \begin{tabular}{|l |c |c |c |c|}
 \hline
 & $B^0,D^0$ & $B^+,D^0$ & $B^0,D^+$ & $B^+,D^+$ \\ [0.5ex]
 \hline \hline
$\Gamma^{\overline{\text{MS}}}_{B_c}$ & 2.97 $\pm$ 0.42 & 2.98 $\pm$ 0.40 & 3.19 $\pm$ 0.80 & 3.19 $\pm$ 0.82 \\
 \hline
 $\Gamma^{\text{meson}}_{B_c}$& 3.03 $\pm$ 0.54 & 3.04 $\pm$ 0.54 & 3.38 $\pm$ 0.98 & 3.39 $\pm$ 0.99 \\
 \hline
 \end{tabular}
 \caption{\small
Here we report the obtained results for the decay rate $\Gamma(B_c)$ in $\text{ps}^{-1}$ from Eq.~\eqref{eq:relBc} in the meson and $\overline{\text{MS}}$ scheme, using the different combinations of $B$ and $D$ mesons.
}
  \label{tab:res}
\end{table}
The results of our approach are shown in Tab.~\ref{tab:res} for the different mesons involved in the relation given in Eq.~\eqref{eq:relBc}. The corresponding uncertainties will be discussed in the discussion section. Note that in the absence of n.p.~and WA/PI corrections, we'd obtained $\Gamma(B^0)+\Gamma(D^0)=3.10\;\text{ps}^{-1}$ and  $\Gamma(B^+)+\Gamma(D^+)=1.57\;\text{ps}^{-1}$: the PI correction  reduces the disagreement between the $B^0D^0$ and $B^+D^+$ estimates of the $B_c$ lifetime from 50\% to 10\%.

\section{Discussion}
\label{Sec:Discussion}
There are a number of implicit assumptions and limitations in the above analysis. We have already mentioned the implicit use of quark hadron duality for the computation of decay widths. The theory of semileptonic decay widths of heavy hadrons uses  an OPE at large complex energy, and relates this to the physical region through a dispersion relation \cite{Chay:1990da}; thus, for the semileptonic width the use of duality rests on the same foundation as, say, the classic computation of the cross section of $e^+e^-\to$~hadrons. On the other hand the theory of inclusive nonleptonic withds   rests on less certain grounds. In this context,  according to Ref.~\cite{Blok:1994cd}  it may well be that the violations to duality fall off as $1/m_Q$ as is the case in the model of Ref.~\cite{Shifman:1994py}; see also Refs.~\cite{Grinstein:1997xk,Grinstein:1998gc,Grinstein:2001nu,Grinstein:2001zq}. It may also be that violations to duality fall off as $\text{exp}(-(m_Q/E_0)^\gamma)$, for some characteristic hadronic scale $E_0$, with $\gamma$ the power that characterises the high order behavior of the expansion in powers of $1/m_Q$ \cite{Blok:1994cd}.

\begin{figure}[t]
    \centering
    \includegraphics[trim={0.3cm 0.3cm 1cm 0}, totalheight=0.25\textheight]{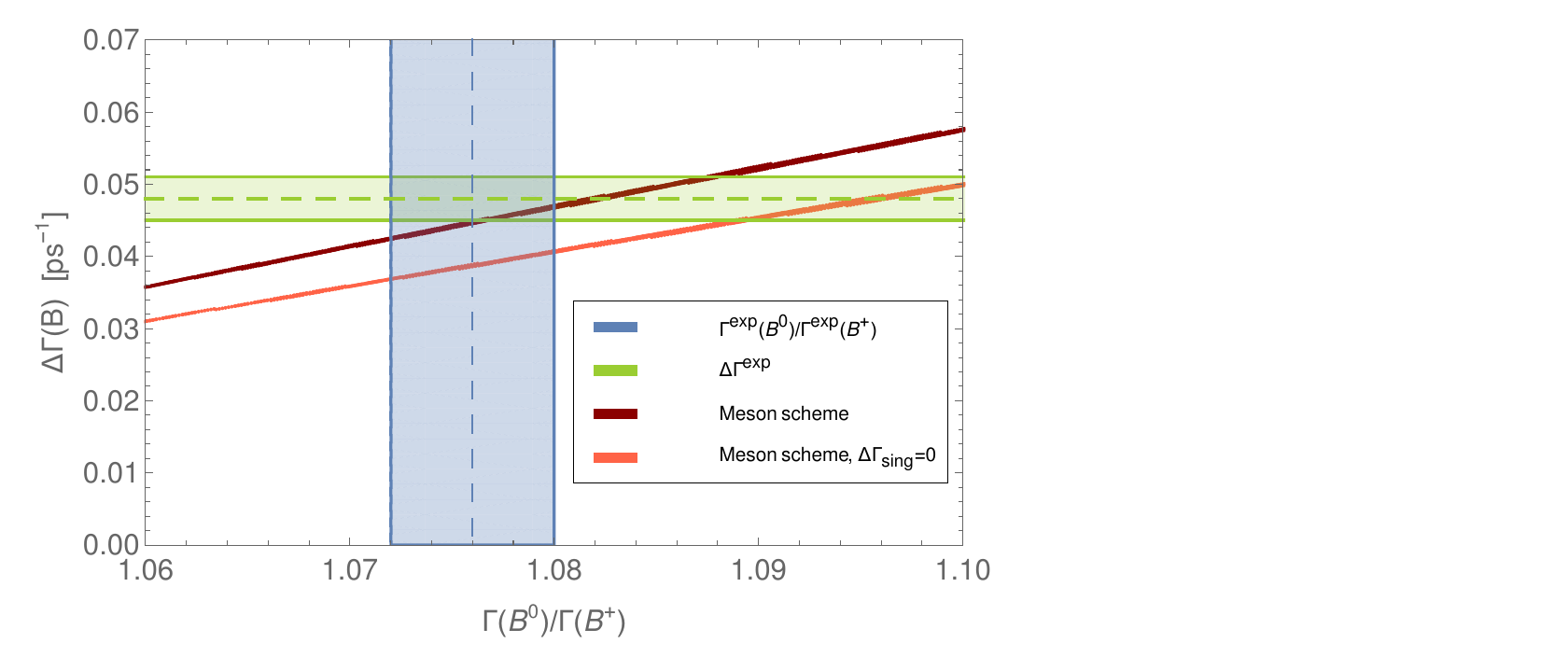}
    \caption{$\Delta\Gamma(B)=\Gamma(B^0)-\Gamma(B^+)$ (in ps$^{-1}$)  vs $r(B)=\Gamma(B^0)/\Gamma(B^+)$ scatter plot, from allowing the bag parameters and the decay constant $f_B$ to vary within their 3$\sigma$ uncertainty: the orange curve shows the result of our calculation and the dark red curve shows the result after adjusting by adding a common, a.k.a. ``singlet'',  contribution, $\Delta\Gamma_{\rm sing}=0.083~\text{ps}^{-1}$, to both $\Gamma(B^0)$ and $\Gamma(B^+)$. }
    \label{fig:rVsDeltaB}
\end{figure}

The $1/m_Q$ expansion seems to work well for $B$ decays. In units of $\text{ps}^{-1}$ the ``singlet width" $\Gamma_{\rm sing}$, defined to be the width exclusive of WA and PI contributions is
\begin{multline}\label{eq:WAPI}
0.546=0.560_{\text{pert}}-0.004_{\lambda_1}-0.012_{\lambda_2}\\
+0.002_{\rho_1}+0.001_{\rho_2}+0.000_{\tau_1}-0.000_{\tau_2}-0.002_{\tau_3}\,,
\end{multline}
where the subscript indicates the term in the expansion. Here and in the following we will only show the values obtained in the meson scheme and refrain from showing the corresponding $\overline{\text{MS}}$ numbers, since the nature of the expansion is the same in both schemes. Because we merely intend to demonstrate the nature of the expansion, only central values are given.  The uncertainty is dominated by that of the perturbative contribution,\footnote{The perturbative calculation is carried out to next to leading order; see \cite{Aebischer:2021ilm} for details.}  which however does not contribute to our central relation, Eq.~\eqref{eq:relBc}. By including the  contributions of WA and PI,  it becomes apparent that $\Gamma^{\text{sing}}(B)$ is underestimated. While attention is often focused on the ratio of widths $r(B)=\Gamma(B^0)/\Gamma(B^+)$ because some uncertainties cancel in the ratio, for our purposes we need values of widths, so in addition to $r(B)$ we investigate the difference $\Delta\Gamma(B)=\Gamma(B^0)-\Gamma(B^+)$.
While we obtain  the central value   $r(B)=1.11$, to be compared with the experimental $r(B)_{\rm PDG}=1.08$, our calculation of the width difference $\Delta\Gamma(B)=0.053\pm0.014\;\text{ps}^{-1}$ slightly overestimates the experimental value of $0.047\pm0.002\;\text{ps}^{-1}$. This is illustrated in Fig.~\ref{fig:rVsDeltaB} that also  shows the effect of uncertainties. The orange curve shows the range of values in our calculation of $r(B)$ and $\Delta\Gamma(B)$ spanned by allowing the decay constant $f_B$ and the bag parameters to vary within their 3$\sigma$ uncertainty, and the dark red curve shows the effect of adding to the previous a singlet contribution $\Delta\Gamma_{\rm sing}=0.083~\text{ps}^{-1}$, chosen to have as many as possible of the scatter plot points within the 1$\sigma$ intersection region of the experimental values for $r(B)$ and $\Delta\Gamma(B)$. Possible origins of $\Delta\Gamma_{\rm sing}$ will be discussed below, but for now
it is interesting to note that the calculation of the inclusive semileptonic width is in good agreement with experiment. In units of $\text{fsec}^{-1}$, we calculate (central value only)
\begin{multline}
\label{eq:BwidthPieces}
    67.0=70.6_{\text{pert}}-0.6_{\lambda_1}-2.4_{\lambda_2}\\
    -0.5_{\rho_1}+0.1_{\rho_2}+0.1_{\tau_1}-0.0_{\tau_2}-0.3_{\tau_3}\,,
\end{multline}
to be compared with the experimental values of $68.0\pm 1.9$ and $67.1\pm 1.7$ for $\Gamma(B^0)$ and $\Gamma(B^+)$, respectively. We have included in Eq.~\eqref{eq:BwidthPieces} the NNLO contribution to the the perturbative contribution \cite{Ciuchini:2001vx,Pak:2008cp}, which is not available for the nonleptonic width. Still, the 1S expansion indicates good convergence of the expansion: in units of the tree-level contribution the nature of the expansion is given by
\[
1-0.11\epsilon-0.03\epsilon^2
\]
where $\epsilon=1$ tracks the 1S expansion parameter \cite{Hoang:1998hm}. This suggests that the underestimate of the total width is due to the nonleptonic contribution, for example, from  violation of duality.\footnote{The leptonic width is used to determine the value of $|V_{cb}|$, and one may argue that the above reasoning is circular. But the value of $|V_{cb}|$ can be independently  obtained from exclusive decays \cite{FermilabLattice:2014ysv} or indirectly from the unitarity triangle \cite{ckmfitter}, and the conclusion remains that the leptonic contribution is in good agreement with experiment. }

\begin{figure}[t]
    \centering
    \includegraphics[trim={0.3cm 0.3cm 1cm 0}, totalheight=0.25\textheight]{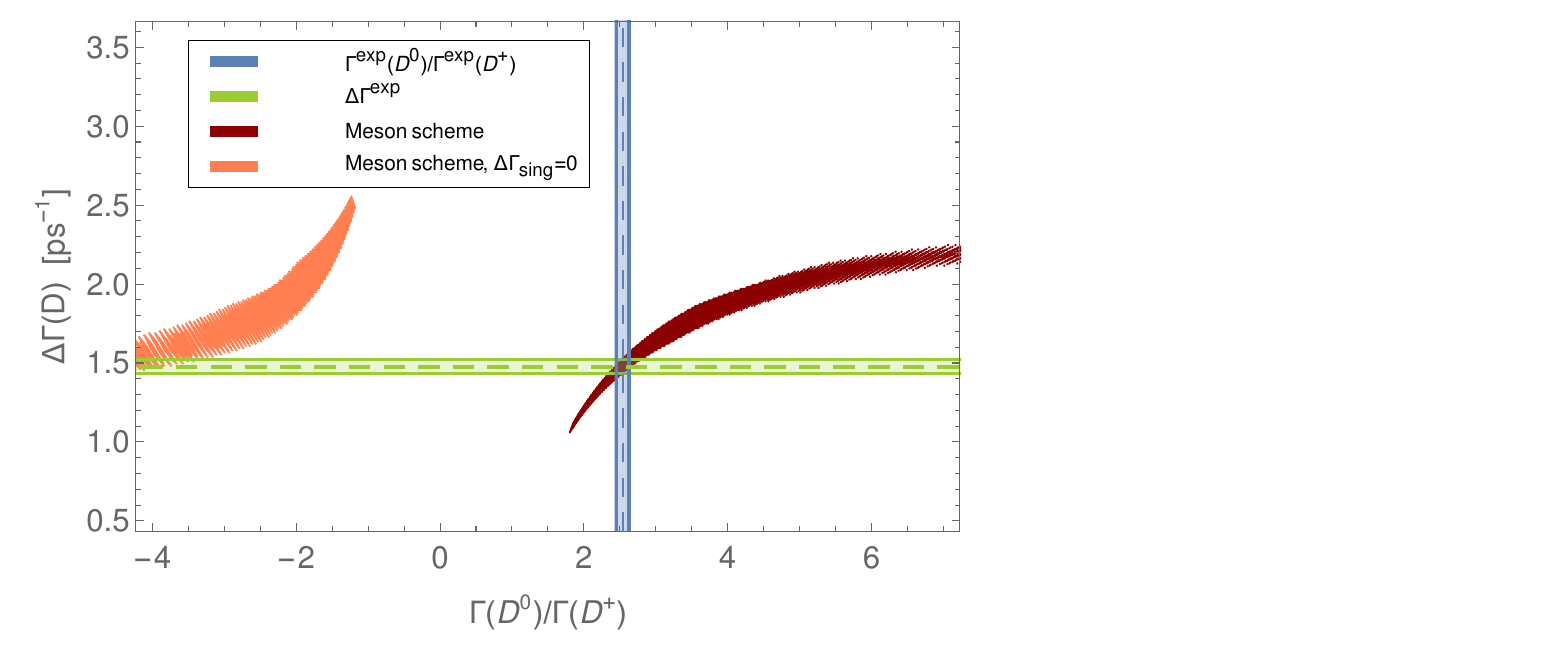}
    \caption{$\Delta\Gamma(D)=\Gamma(D^0)-\Gamma(D^+)$ (in ps$^{-1}$)  vs $r(D)=\Gamma(D^0)/\Gamma(D^+)$ scatter plot, from allowing the bag parameters and the decay constant $f_D$ to vary within their 3$\sigma$ uncertainty: the orange curve shows the result of our calculation and the dark red curve shows the result after adjusting by adding a common, a.k.a. ``singlet'',  contribution, $\Delta\Gamma_{\rm sing}=1.2~\text{ps}^{-1}$, to both $\Gamma(D^0)$ and $\Gamma(D^+)$. }
    \label{fig:rVsDeltaD}
\end{figure}

The situation is less clear for the case of $D$ decays, because of the lower scale of the charm mass. The breakdown of contributions to the width, exclusive of WA and PI terms, in analogy with Eq.~\eqref{eq:BwidthPieces}, is
\begin{multline}
\label{eq:DratePieces}
1.187=1.168_{\text{pert}}-0.086_{\lambda_1}-0.040_{\lambda_2}
+0.137_{\rho_1}\\
+0.008_{\rho_2}+0.029_{\tau_1}-0.009_{\tau_2}-0.019_{\tau_3}\,.
\end{multline}
This however cannot be compared immediately with experiment since the PI and WA contributions are quite significant, and are expected to account for the large difference in $D^+$ and $D^0$ lifetimes \cite{Guberina:1979xw}, $r(D)=\Gamma(D^0)/\Gamma(D^+)=2.536\pm0.026$ \cite{Tanabashi:2018oca}.
Fig.~\ref{fig:rVsDeltaD}  shows the result once WA and PI contributions are included. The orange curve shows the range of values in our calculation of $r(D)$ and $\Delta\Gamma(D)$ spanned by allowing the decay constant $f_D$ and the bag parameters to vary within their 3$\sigma$ uncertainty, and the dark red curve shows the effect of adding to the previous a singlet contribution $\Delta\Gamma_{\rm sing}=1.2~\text{ps}^{-1}$, chosen to have as many as possible of the scatter plot points within the 1$\sigma$ intersection region of the experimental values for $r(D)$ and $\Delta(D)$. Before discussing the origin of $\Delta\Gamma_{\rm sing}$ it is instructive to inspect the semileptonic decay width, that is predicted well in the case of $B$ mesons, as discussed above. In units of ps$^{-1}$  we obtain, using $\mu=m_c$ for the renormalization point, and including only next to leading order corrections,
\begin{multline}
\label{eq:DsemilepPieces}
0.121=0.166_{\text{pert}}-0.015_{\lambda_1}-0.042_{\lambda_2}
+0.020_{\rho_1}\\
+0.008_{\rho_1}+0.005_{\tau_1}-0.002_{\tau_2}-0.020_{\tau_3}\,.
\end{multline}
somewhat shy of the PDG values  0.158~ps$^{-1}$ and 0.155~ps$^{-1}$ for $D^0$ and $D^+$, respectively. However the perturbative expansion indicates slow convergence: using $\mu=m_c$ for the renormalization point, we find
\[
1+0.29\epsilon+0.21\epsilon^2\,.
\]
for the nature of the expansion. This is as expected from the larger value of the strong coupling constant at this lower energy scale. In fact, using $\mu=2m_c$ we find a semileptonic width of 0.156~ps$^{-1}$, with an expansion that exhibits the poor convergence more clearly, $1+0.10\epsilon+0.26\epsilon^2$, suggesting an uncertainty of order 10\% in the NNLO perturbative calculation.

We have used only the meson scheme for $B$ and $D$ decays. It has been remarked that for the case of  $D$ decays  the 1S scheme for the $c$-quark gives an apparently faster convergence of the free quark semileptonic decay rate and that the meson semileptonic width is in good agreement with experiment once a large $1/m_c^2$ correction is included. Including now the $1/m^3$ corrections we find, in analogy to \eqref{eq:DratePieces} but for the 1S scheme
\begin{multline}
1.933=1.967_{\text{pert}}-0.013_{\lambda_1}-0.086_{\lambda_2}
+0.178_{\rho_1}\\
+0.015_{\rho_2}+0.038_{\tau_1}-0.012_{\tau_2}-0.035_{\tau_3}\,.
\end{multline}
Both this expansion, and the one in \eqref{eq:DratePieces} perhaps indicate poor convergence of the $1/m$ expansion. Still, the overall non-perturbative correction is small in both schemes. The inclusive semileptonic width, which seems to give a very good agreement with experiment in the 1S scheme when $1/m^2$ corrections are included is now (in analogy with \eqref{eq:DsemilepPieces})
\begin{multline}
0.196=0.268_{\text{pert}}-0.024_{\lambda_1}-0.065_{\lambda_2}
+0.027_{\rho_1}\\
+0.011_{\rho_2}+0.007_{\tau_1}-0.002_{\tau_2}-0.027_{\tau_3}\,.
\end{multline}
The perturbative expansion is much better behaved in this scheme:
\[
1-0.122\epsilon+0.013\epsilon^2\,.
\]
Had we included the NNLO perturbative contribution, we'd obtain $0.202~\text{ps}^{-1}$ for the semileptonic width. The excellent agreement with the measured width noted in Ref.~\cite{Hoang:1998ng} is diminished by the inclusion of $1/m^3$ nonperturbative corrections.  That work points out the presence of large order $\epsilon^3$ corrections from the expansion of the $J/\psi$ mass; however, the nature of the expansion cannot be ascertained without a full computation of the NNNLO c-quark semileptonic decay width.

The need for a  singlet contribution to account for the $B$ and $D$ widths may be the result of a violation to duality. However, duality is not in question for inclusive semileptonic widths. While the semileptonic width for $B$ decays is in fair agreement with data, the one for $D$ decays exhibits  tension with data. As seen above, the meson scheme underestimates the rate while the 1S scheme overestimates it. This suggests that this may well be the result of a poorly convergent perturbative expansion.   Assuming duality is not violated for hadronic decays, the power of our proposal becomes apparent: the slow convergence of the perturbative expansion, which is the leading contribution to the rates, is immaterial. And while the $1/m$ expansion also converges rather slowly, its magnitude is comparatively small, therefore leading to a better controlled estimate of the $B_c$ width by use of Eq.~\eqref{eq:relBc}.

To sharpen this observation, we naively consider the possibility that duality violation in the nonleptonic (n.l.) rate scales as a power of the heavy quark mass. If we assume the ratio $(\Delta\Gamma_{\rm sing}^{\rm n.l.}(B)/|V_{cb}|^2)/(\Delta\Gamma_{\rm sing}^{n.l.}(D)/|V_{cs}|^2)\sim 45$ scales as $(m_b/m_c)^{5-p}$ for some power $p$,  we find $p\lesssim 2$. Alternatively, if $\Delta\Gamma_{\rm sing}$ is largely due  to next order corrections, we would expect $\Delta\Gamma_{\rm sing}(B)/\Delta\Gamma_{\rm sing}(D)\sim (m_b/m_c)^5 (\alpha_s(m_b)/\alpha_s(m_c))^2$ which is of order $10^2$. Pending a calculation of the NNLO perturbative hadronic decay, we assume duality holds for these decays. Moreover, it should be kept in mind that there are uncertainties in the  estimate of $\Delta\Gamma_{\rm sing}$ from $1/m^4$ 4-quark operators that contribute to WA and PI. In addition, even at order $1/m^3$ the 4-quark operators contribute to  $\Delta\Gamma_{\rm sing}$ through the ``eye-graphs" in the non-perturbative calculation of matrix elements that, as mentioned above, have not been computed. While these eye-graphs are naively suppressed by a loop factor relative to the graphs retained in the calculation of bag parameters, the naive expectation of suppression of these graphs may be wrong; see Ref.~\cite{Manohar:2013rga} for an example of a model in which such a loop suppression is absent in a calculable toy model. Since the coefficients of the 4-quark operators  have a phase space enhancement factor of $\sim16\pi^2$ and are suppressed by $1/m_b^3$ for $B$ decays, which one may estimate as $(\sqrt{-\lambda_1}/m_b)^3\approx0.2\%$, the  contribution for the eye-graph, if not suppressed by the internal loop, could be as large as $~16\pi^2(0.002)\sim 30\%$, which is sufficient by itself to account for $\Delta\Gamma_{\rm sing}(B)$. And although we argued above for a scaling with $p\lesssim2$ while the eye-graph contribution scales as $p=3$, our scaling estimate is itself subject to  uncertainties as seen in Fig.~\ref{fig:rVsDeltaB} (and, although less apparent, also in Fig.~\ref{fig:rVsDeltaD}).

\begin{figure}[b]
    \centering
    \includegraphics[trim={0.3cm 0.3cm 1cm 0}, totalheight=0.25\textheight]{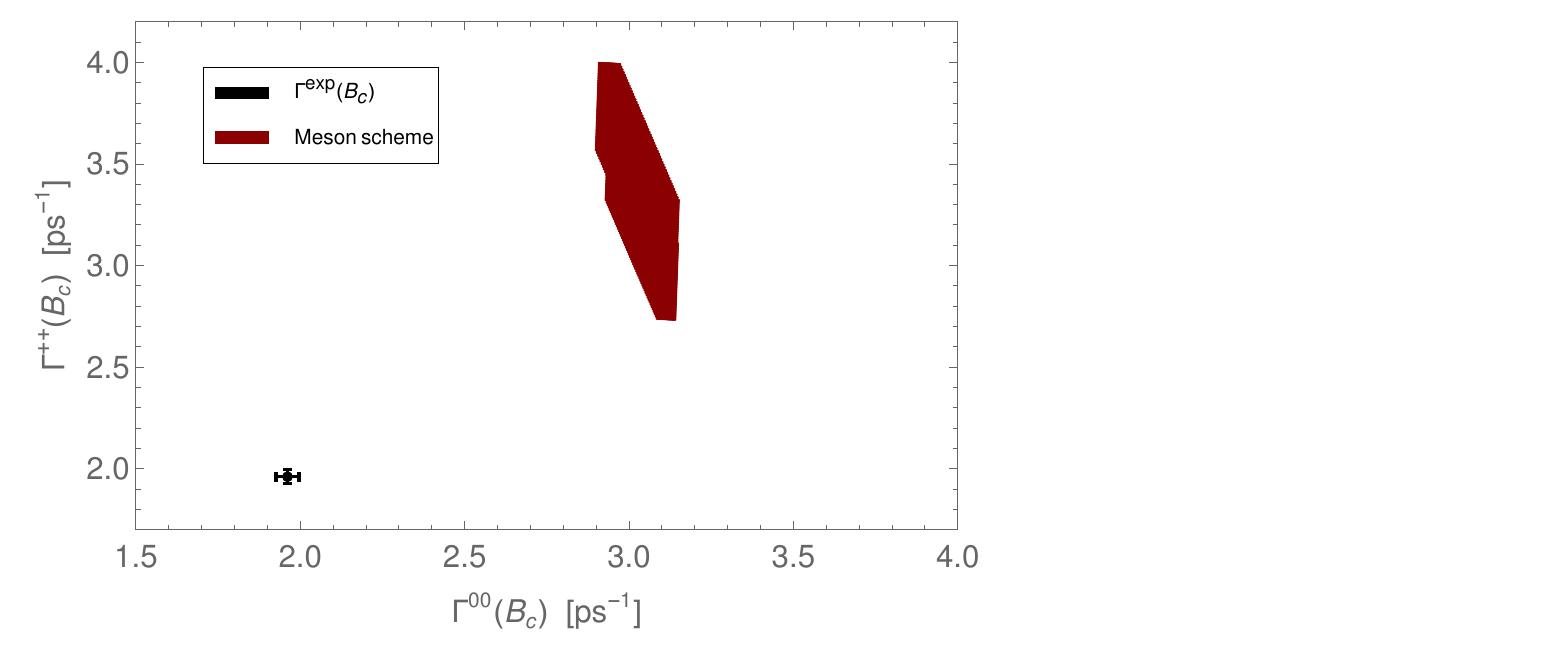}
    \caption{$\Gamma^{00}(B_c)$  vs $\Gamma^{++}(B_c)$ (in ps$^{-1}$) scatter plot, from allowing the bag parameters and the decay constants to vary within their 3$\sigma$ uncertainty. $\Gamma^{00}(B_c)$ is obtained from the $B^0$ and $D^0$ modes whereas $\Gamma^{++}(B_c)$ is computed using $B^+$ and $D^+$. The dark red area shows the result of our calculation and the black point represents the experimental value. }
    \label{fig:gamma00pp}
\end{figure}

The computation of the width for the $B_c$ meson is based on NRQCD, that is, an expansion in velocity, $v$, rather than in $1/m_Q$. In units of $\text{ps}^{-1}$ the ``singlet width" $\Gamma_{\rm sing}$ is
\begin{align*}
1.628&=1.728_{\rm pert}-0.100_{\rm n.p.} & &{\rm (meson)}\\
2.322&=2.527_{\rm pert}-0.205_{\rm n.p.} & &{\rm (1S)}
\end{align*}
The non-perturbative contributions are dominated by $c$-quark decay. Hence,
their  magnitude is slightly smaller than  expected,   $v^2\sim\alpha_s(m_c)^2\sim0.14$. The clear dominance of the  perturbative term indicates, that it is likely to produce the leading uncertainty. This observation, of course, is the motivation for the present work; we remind the reader: calculating the width using Eq.~\eqref{eq:relBc} eliminates the perturbative contribution from the calculation.  The width computed in Ref.~\cite{Aebischer:2021ilm} in both meson ad 1S schemes, $\Gamma^{\rm meson}_{B_c}=(1.70\pm0.46)\;\text{ps}^{-1}$ and $\Gamma^{\rm 1S}_{B_c}=(2.40\pm0.42)\;\text{ps}^{-1}$,  is consistent with the experimental one; see below, Eq.~\eqref{eq:BcWidthPDG}.  The uncertainty is split about equally between NNLO corrections to perturbative,  and non-perturbative corrections. However, the leading contribution to the non-perturbative uncertainty is from the coefficient of the order $v^2$ (leading) operator, which is only known to LO, which could be computed when necessary.

Finally, we discuss the results obtained from our approach, which are summarized in Tab.~\ref{tab:res}. The  values we obtain for the $B_c$ decay rate differ quite significantly from the experimental value \cite{ParticleDataGroup:2014cgo}
\begin{equation}
\label{eq:BcWidthPDG}
\Gamma^{\text{exp}}_{B_c}=1.961(35) \text{ps}^{-1}\,.
\end{equation}
This  is  illustrated in Fig.~\ref{fig:gamma00pp}, where the black cross shows the experimental values and the red area shows the values of computed decay rates after variation of the bag parameters and decay constants within their 3$\sigma$ ranges. This result is unchanged when bag parameters from HQET sum rules \cite{Kirk:2017juj}  are used, since they lie within the 3$\sigma$ range of the ones from \cite{Becirevic:2001fy}.
To illustrate the uncertainties in Tab.~\ref{tab:res} we discuss the uncertainty in the first entry, $\Gamma^{00}(B_c)$, that is, for the decay  of  the neutral mesons, $B^0$ and $D^0$.
The uncertainty in $\Gamma^{00}(B_c)$ is given by
\begin{multline}\label{eq:error}
    0.54=0.20_{\Gamma^{\rm n.p.}( B_c)}+0.12_{\lambda_1,\rho_1,\tau_{1,2,3}}+0.13_{\Gamma^{\text{WA}}(D^0)}\\
    +0.08_{\Gamma^{\rm n.p.}(D^0)}+0.01_{\text{exp}}\,.
\end{multline}
The leading uncertainty results from n.p. corrections to the $B_c$, which are discussed in detail in \cite{Aebischer:2021ilm}. The second dominant uncertainty stems from the n.p. parameters $\lambda_1,\rho_1,\tau_{1,2,3}$, which we estimate using the correlation matrix in Tab.~80 of \cite{HFLAV:2019otj} for the error propagation. The uncertainty from $\Gamma^{\text{WA}}(D^0)$ contains $1/m$ as well as NNLO QCD corrections to the WA contributions, as well as the uncertainties from the bag parameters. Estimates of the $1/m_c^4$ corrections are included in the uncertainty $\Gamma^{\rm n.p.}(D^0)$ and the last uncertainty is due to experimental uncertainties. The much larger uncertainties in Tab.\ref{tab:res} result from the sizable $\Gamma^{\rm PI}(D^+)$ contributions, compared to the much smaller $\Gamma^{\rm WA}(D^0)$ contributions.

\section{Conclusions}
\label{Sec:Conclusions}
We have proposed a novel way to determine the $B_c$ lifetime from differences of $B_c$, and  $B$ and $D$ lifetimes. Following this approach we find values for the $B_c$ decay rate, which lie significantly above the experimental value; see Tab.~\ref{tab:res}. The  uncertainties of our approach are summarized in Eq.~\eqref{eq:error} for the case of neutral $B$ and $D$ mesons. The largest uncertainty is from the non-perturbative correction to the $B_c$ decay width, and is dominated by the NLO correction to the coefficient of the order $v^2$ operator in the NRQCD expansion. The parametric uncertainty from the non-perturbative parameters of the $1/m$ expansion, $\lambda_1,\rho_1,\tau_{1,2,3}$, is likely to be reduced with future data. The remaining substantial uncertainties are from NNLO corrections and from $1/m^4$ effects, both to the WA and PI as well as to the iso-singlet contributions.  This motivates to investigate the higher order corrections to these processes, as done for semileptonic decays in \cite{Dassinger:2006md}.

The most conservative explanation of the tension between our results and the experimental width is that we have underestimated  uncertainties. Independent verification, not just of our calculation but of inputs, like bag parameters, that have not been independently verified, is of course necessary to lend confidence to the claim of tension. One possible source of a large correction is the omitted eye-graphs in the non-perturbative calculation of the matrix elements of 4-fermion operators. A recent study however suggests that these contributions are not sizable \cite{King:2021jsq}. There are only two other possible explanations we can offer, considering that corrections of dimension seven operators have been estimated in our calculation. The more radical one, that New Physics is at play, is somewhat peculiar in that the New Physics would have to interfere destructively with the SM amplitude. Perhaps a more conservative explanation is that quark hadron duality does not hold for non-leptonic decays \cite{Blok:1994cd}. In our discussion we have noted that in order to reproduce the widths of $D$ and $B$ mesons we found it necessary to postulate  the addition of an iso-singlet correction $\Delta\Gamma_{\rm sing}$; this is illustrated in Figs.~\ref{fig:rVsDeltaB} and~\ref{fig:rVsDeltaD}. We proceeded above under the assumption that these corrections can be attributed to an underestimate of the singlet contribution to the width, most likely from  an unexpectedly large NNLO perturbative correction to the quark decay rate. But, by design, this cancels out in our method's calculation of $\Gamma(B_c)$. If, on the other hand, $\Delta\Gamma_{\rm sing}(B)$ and $\Delta\Gamma_{\rm sing}(D)$ are due to duality violation in the hadronic decay widths, then there is no reason a priori that one expects them to cancel against a duality violating contribution, $\Delta\Gamma_{\rm sing}(B_c)$,  to the $B_c$ decay width. The very nice agreement of the calculation of the $B_c$ life-time with the experimental value in Ref.~\cite{Aebischer:2021ilm} suggests that duality violation is small in that case. The case for duality violation in $B$ and $D$, but but not in $B_c$,  decays is strengthened by the observation that the  differences  $\Gamma(B_c)^{00}-\Gamma(B_c)_{\rm PDG}\approx\Gamma(B_c)^{+0}-\Gamma(B_c)_{\rm PDG}\approx(1.1\pm0.5)\;\text{ps}^{-1}$ and $\Gamma(B_c)^{0+}-\Gamma(B_c)_{\rm PDG}\approx\Gamma(B_c)^{++}-\Gamma(B_c)_{\rm PDG}\approx(1.4\pm1.3)\;\text{ps}^{-1}$   all agree with the excess inferred solely from consideration of $B$ and $D$ meson lifetimes:  $\Delta\Gamma_{\rm sing}(B)+\Delta\Gamma_{\rm sing}(D)=1.3\;\text{ps}^{-1}$.

\smallskip

{\it Acknowledgments} --- {J.A.\ acknowledges financial support from the Swiss National Science Foundation (Project No. P400P2\_183838). The work
  of B.G.\ is supported in part by the U.S. Department of Energy Grant No. DE-SC0009919. }

  \bibliography{AGBIB}

\end{document}